\documentclass[11pt,twoside]{article}

\usepackage{asp2014}

\aspSuppressVolSlug
\resetcounters

\begin{document}

\title{Prototype Implementation of a Web-Based Gravitational Wave Signal Analyzer: SNEGRAF}

\author{Satoshi~Eguchi,$^1$ Shota~Shibagaki,$^1$ Kazuhiro Hayama,$^1$ and Kei Kotake$^1$}
\affil{$^1$Fukuoka University, 8-19-1, Nanakuma, Jonan-ku, Fukuoka 814-0180, Japan; \email{satoshieguchi@fukuoka-u.ac.jp}}

\paperauthor{Satoshi~Eguchi}{satoshieguchi@fukuoka-u.ac.jp}{0000-0003-2814-9336}{Fukuoka University}{Department of Applied Physics, Faculty of Science}{Fukuoka City}{Fukuoka Prefecture}{814-0180}{Japan}



\begin{abstract}
A direct detection of gravitational waves is
one of the most exciting frontiers for modern
astronomy and astrophysics.
Gravitational wave signals combined with classical
electro-magnetic observations, known as multi-messenger astronomy,
promise newer and deeper insights about
the cosmic evolution of astrophysical objects
such as neutron starts and black holes.
To this end, we have been developing an original
data processing pipeline for KAGRA,
a Japanese gravitational wave telescope,
for optimal detections of supernova events.
As a part of our project, we released a web
application named SuperNova Event Gravitational-wave-display
in Fukuoka (SNEGRAF) in autumn 2018.
SNEGRAF accepts the users' theoretical waveforms
as a plain text file consisting of a time series
of $h_{+}$ and $h_{\times}$ (the plus and cross
mode of gravitational waves, respectively),
then displays the input, a corresponding spectrogram,
and power spectrum together with KAGRA sensitivity curve
and the signal-to-noise ratio;
we adopt Google Visualization API for the interactive
visualization of the input waveforms.
However, it is a time-consuming task to draw
more than $\sim 10^{5}$ data points directly
with JavaScript, although the number can be
typical for a supernova hunt by assuming a typical
duration of the event and sampling rate of the detectors;
a combination of recursive decimations of
the original in the server-side program and
an appropriate selection of them depending
on the time duration requested by the user
in a web browser achieves an acceptable latency.
In this paper, we present the current design,
implementation and optimization algorithms of SNEGRAF,
and its future perspectives.
\end{abstract}


\section{Introduction}

In the framework of general relativity,
a mass curves the space-time around it,
and the curvature is observed as gravity.
An accelerated motion of a mass generates a disturbance
of space-time, which propagates in a vacuum in the form
of waves;
these waves are referred to as ``gravitational waves.''
Gravitational waves can penetrate even a very dense material,
and carry the information of the space-time around a massive
but compact astronomical object such as a neutron star and
black hole.
The first direct detection of a gravitational wave is
known as GW150914, where a merger of two stellar-mass
black holes took place \citep{2016PhRvL.116f1102A}.

Multi-messenger astronomy, which utilizes observations
of gravitational waves and neutrinos combined with
those in multiple wavelengths, attracts a lot of attention
recently since it promises a deeper understanding of
the innermost part of a high energy astrophysical phenomenon.
For example, roughly two explosion mechanisms are proposed
for a core-collapse supernova to date, leading to outstanding
differences in their gravitational waveforms
\citep[see][and references therein]{2013CRPhy..14..318K}.
Hence, at Fukuoka University, we assembled a team to promote
multi-messenger astronomy focusing on the physics of
supernovae in April 2018.
Goals of our mission are:
\begin{itemize}
	\item Developing an original data processing
	pipeline for KAGRA, a Japanese gravitational
	wave telescope, to detect a supernova event
	at optimal efficiency,
	\item Providing data visualization and analysis
	software for the KAGRA observations to the world.
\end{itemize}

\section{SNEGRAF}

\begin{figure}
	\plotone{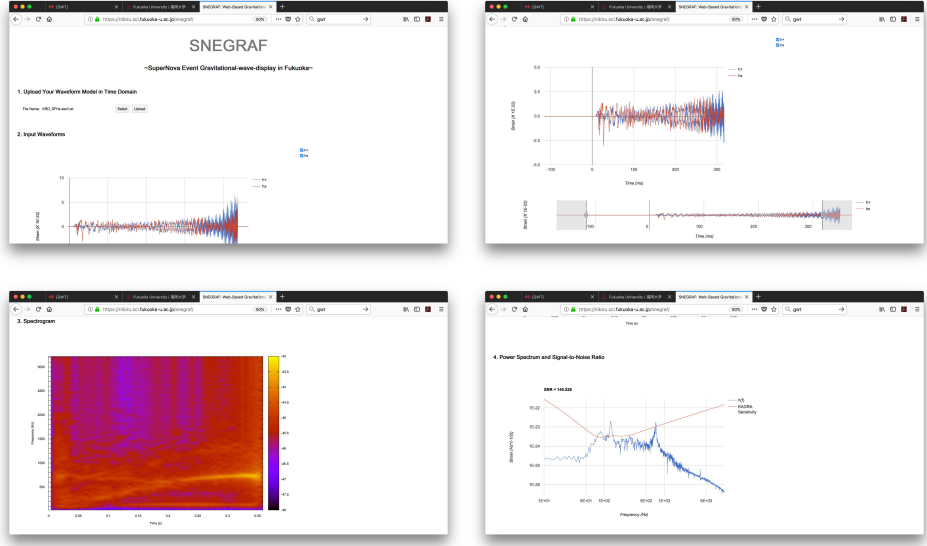}
	\caption{Screenshots of SNEGRAF.
		From left to right and top to bottom,
		the banner, waveform viewer, spectrogram,
		and power spectrum, respectively.\label{P8-2-fig-snegraf-screenshot}}
\end{figure}

As the first step of our software releases,
we have just made a web application named
``SuperNova Event Gravitational-wave-display in Fukuoka
(SNEGRAF; Fig.~\ref{P8-2-fig-snegraf-screenshot})''
public in October 2018.
SNEGRAF accepts a time series of $h_{+}$ and $h_{\times}$
(two individual modes of a gravitational wave) in a character-separated-value (CSV)
format as an input (Table~\ref{P8-2-tab-format}), and displays the input waveforms, a corresponding
spectrogram, and power spectrum together with the signal-to-noise ratio
of the input signal and the analytic KAGRA sensitivity curve.
The access url to SNEGRAF is \url{https://nibiru.sci.fukuoka-u.ac.jp/snegraf/}.

\begin{table}
	\centering
	\begin{tabular}{cccc}
		\hline
		Column Number & 1st & 2nd & 3rd \\
		\hline
		Content & Time (sec) & $h_{+}$ & $h_{\times}$ \\
		\hline
	\end{tabular}
	\caption{Details of an input file format for SNEGRAF.
		A pipe (\texttt{|}), comma (\texttt{,}), tab (\texttt{\symbol{"5C}t}),
		and white space are acceptable for a column separator.
		A hash (\texttt{\#}) is regarded as a beginning of comments. \label{P8-2-tab-format}}
\end{table}

SNEGRAF is a simple Ajax application hosted on a Java servlet.
Since we have quite limited human resources and utilize
existing software libraries written in either C/C++ or Java,
we adopt GWT\footnote{It was known as Google Web Toolkit previously. \url{http://www.gwtproject.org/}},
which generates both server-side and client-side codes
from a single Java source file, for an application framework.
Google Charts\footnote{It is also known as Google Visualization API. \url{https://developers.google.com/chart/}}
and its GWT binding\footnote{GWT Charts. \url{https://github.com/google/gwt-charts/}}
are used for an interactive visualization
of input waveforms.

To simplify the server-side programs,
the file uploading functionality is implemented with
File API in HTML5.
A text file uploaded by a user is transfered to the servlet
as is as an argument of type \texttt{String} during a remote
procedure call (RPC).
Then the string is parsed into arrays of type \texttt{double}
to hold $(t, h_{+}, h_{\times})$ in each row in the servlet,
and ``resampled and decimated hierarchically (see Sect.~\ref{P8-2-sec-data-reduction}).''
The servlet invokes a Python script to compute a spectrogram,
which is converted to a scalable vector graphics (SVG) file
by \texttt{gnuplot} and encoded into a Base64 string.
A power spectrum is calculated with a Java implementation
of fast Fourier transform (FFT)\footnote{\url{https://www.nayuki.io/page/free-small-fft-in-multiple-languages}},
accompanied by an evaluation of the signal-to-noise ratio (SNR)
based on the analytic KAGRA sensitivity curve \citep{2012arXiv1202.4031M}.
Note that detector beam-pattern functions are assumed to be
unity in the estimation.
At the end, the waveform arrays, the Base64 encoded spectrogram,
the array for the power spectrum, and the SNR are packed into
a single object in JavaScript object notation (JSON), and
returned to the web client as the result of the RPC.
The waveforms and power spectrum are plotted with Google
Charts on the client.

\section{Data Reduction Algorithm} \label{P8-2-sec-data-reduction}

By assuming a typical sampling rate of a gravitational
wave detector, our programs should be able to handle
$N \sim 10^{5}$ data points on the fly.
However, this is a very heavy task for a JavaScript application
like SNEGRAF currently.
To achieve this goal even on a low-powered CPU,
we applied ``hierarchical decimation technique''
to SNEGRAF.
The basic idea of this method is to apply a decimation by
a factor of 2 recursively on server side, and to select
an adequate result depending on the time duration requested
by a user on client side.

\begin{enumerate}
	\item Find the integer $m$ satisfying $2^{m} < N \leq 2^{m + 1}$
	and resample the original waveform evenly into new $2^{m}$ points by linear interpolation.
	This takes $O \left( m 2^{m} \right)$ time.
	\item Decimate the resampled waveform by 2.
	This yields new $N_{m - 1} = 2^{m - 1}$ data points and takes $O \left( 2^{m - 1} \right)$ time.
	\item Apply the 2nd step recursively until
	the number of new data points $N_{i}$ is less than $N_{\text{disp}, \text{th}} \left( = 2048 \right)$.
	At this step, the total amount of data points is exactly less than $N + N / 2 + N / 4 + \cdots = 2 N$.
	\item On client side, calculate the number of data points $N_{\text{disp}, i}$
	which fall inside the user specified time range for each decimated data.
	The total processing time is $O \left( m^{2} \right)$.
	\item On the client, find the largest $i$ such that $N_{\text{disp}, i} \leq N_{\text{disp}, \text{th}}$
	with a binary search algorithm, and plot the data.
\end{enumerate}

When $N$ is $\sim 10^{5} \simeq 2^{17}$, there are just
$\simeq 150$ lookups of the arrays and $\leq N_{\text{disp}, \text{th}}$
drawings on the client with just consuming twice as much as the initial memory space.
The processing time on client side is reduced
by two orders of magnitude thanks to this algorithm,
and SNEGRAF quickly responds to the user's operations
even on a low-powered computer with an Intel Atom CPU.

\section{Future Work}

\begin{itemize}
	\checklistitemize
	\item A spectrogram and power spectrum displayed on
	the current version are ``static.''
	We have a plan to make them interactive
	(e.g., the time range on the input waveform viewer
	will link to that selected on a spectrogram).
	\item To display a sky map, which is a heat map
	representing the likelihood of the source direction.
\end{itemize}

\acknowledgements

This work is supported by JSPS KAKENHI Grant Number 17H06364.

\bibliography{P8-2}

\end{document}